\def\draftversion{false}

\RequirePackage{ifthen}
\ifthenelse{\equal{\draftversion}{true}}{
  \documentclass[prb,galley,showpacs,preprintnumbers,citeautoscript,
      amsmath,amssymb,longbibliography]{revtex4-2}
}{
  \documentclass[citeautoscript,floatfix,aps,prb,twocolumn,
      superscriptaddress,longbibliography]{revtex4-2}
}

\usepackage{amsmath,latexsym,psfrag}
\usepackage[pdftex]{graphicx}

\usepackage{ragged2e} 
\usepackage{subcaption} 

\usepackage{epstopdf}
\usepackage{natbib}
\usepackage{array}
\usepackage{dcolumn}
\usepackage{bm}

\usepackage{soul}  

\usepackage{float}

\usepackage[usenames,dvipsnames]{color}

\soulregister\cite7
\usepackage[%
  colorlinks=true,
  urlcolor=blue,
  linkcolor=blue,
  citecolor=blue
]{hyperref}

\ifthenelse{\equal{\draftversion}{true}}{
  \usepackage{showlabels}
  \marginparwidth 2.7in
  \marginparsep 0.5in
  \newcounter{comm} 
  \def\commnext{\stepcounter{comm}}
  \def\commtext{{\bf\color{blue}[\arabic{comm}]}}
  \def\commmar{{\bf\color{blue}[\arabic{comm}]}}
  \def\msm#1{\commnext\marginpar{\small MS\commmar: #1}\commtext}

  \def\mlab#1{\marginpar{\small\bf #1}}
  
}{
  \def\dvm#1{}
  \def\cdm#1{}
  \def\msm#1{}
  \def\asm#1{}
  \def\miq#1{}
  \def\mlab#1{}
  
}

\begin{document}

\title{Structural and electronic properties of bulk Li$_{2}$O$_{2}$: first-principles simulations based on numerical atomic orbitals}

\author{ Paul M. Masanja }
\affiliation{ Department of Physics, 
              College of Natural and Mathematical Sciences, 
              University of Dodoma, PO Box 259, 
              Dodoma, Tanzania}

\author{ Toraya Fern\'andez-Ruiz }
\affiliation{ Departamento de Ciencias de la Tierra y
              F\'{\i}sica de la Materia Condensada, Universidad de Cantabria,
              Avenida de los Castros s/n, 39005 Santander, Spain}

\author{ Esther J. Tarimo }
\affiliation{ Department of Physics, 
              College of Natural and Mathematical Sciences, 
              University of Dodoma, PO Box 259, 
              Dodoma, Tanzania}

\author{ Nayara Carral-Sainz }
\affiliation{ Departamento de Ciencias de la Tierra y
              F\'{\i}sica de la Materia Condensada, Universidad de Cantabria,
              Avenida de los Castros s/n, 39005 Santander, Spain}

\author{ P.V. Kanaka Rao }
\affiliation{ Department of Physics, 
              College of Natural and Mathematical Sciences, 
              University of Dodoma, PO Box 259, 
              Dodoma, Tanzania}
              
\author{ Vijay Singh }
\affiliation{ Department of Physics, 
              College of Natural and Mathematical Sciences, 
              University of Dodoma, PO Box 259, 
              Dodoma, Tanzania}

\author{ Bernard Mwankemwa }
\affiliation{ Department of Physics, 
              College of Natural and Mathematical Sciences, 
              University of Dodoma, PO Box 259, 
              Dodoma, Tanzania}

\author{ Juan Mar\'{\i}a Garc\'{\i}a-Lastra}
\affiliation{ Department of Energy Conversion and Storage, Technical University of Denmark, Kongens Lyngby DK-2800, Denmark}

\author{ Pablo Garc\'{\i}a-Fern\'andez}
\affiliation{ Departamento de Ciencias de la Tierra y
              F\'{\i}sica de la Materia Condensada, Universidad de Cantabria,
              Avenida de los Castros s/n, 39005 Santander, Spain}

\author{ Javier Junquera }
\affiliation{ Departamento de Ciencias de la Tierra y
              F\'{\i}sica de la Materia Condensada, Universidad de Cantabria,
              Avenida de los Castros s/n, 39005 Santander, Spain}

\date{\today}

\begin{abstract}
  The development of advanced materials with high specific energy is crucial for enabling sustainable energy storage solutions, particularly in applications such as lithium-air batteries. Lithium peroxide (Li$_{2}$O$_{2}$) is a key discharge product in non-aqueous lithium-air systems, where its structural and electronic properties significantly influence battery performance. In this work, we investigate the atomic structure, electronic band structure, and Wannier functions of bulk Li$_{2}$O$_{2}$ using density functional theory. The performance of different basis sets of numerical atomic orbitals are compared with respect to a converged plane-wave basis results. We analyze the material’s ionic characteristics, the formation of molecular orbitals in oxygen dimers, and the band gap discrepancies between various computational approaches. Furthermore, we develop a localized Wannier basis to model electron-vibration interactions and explore their implications for polaron formation. Our findings provide a chemically intuitive framework for understanding electron-lattice coupling and offer a basis for constructing reduced models that accurately describe the dynamics of polarons in Li$_{2}$O$_{2}$. These insights contribute to the broader goal of improving energy storage technologies and advancing the field of materials design.
\end{abstract}

\maketitle

\section{Introduction}
\label{sec:intro}

The development of materials capable of storing energy with minimal weight, i.e., achieving high specific energy, is critically important for the vehicle industry. This objective is a key milestone in the transition away from fossil fuels in the global economy~\cite{Albertus-2011,Radin-13,Tian-14,Bruce-12}. Among the most promising technologies competing with the specific energy of gasoline (46.4 MJ/kg) are lithium/air or lithium/oxygen batteries, which exhibit typical specific energy values of approximately 40 MJ/kg~\cite{Naqvi-22}.

In non-aqueous lithium/air batteries, where the electrolyte is not carbonate-based, the primary discharge product in the cathode is lithium peroxide (Li$_2$O$_2$)~\cite{Sousa-24}. This compound, commonly identified in batteries via Raman spectroscopy~\cite{Sousa-24}, is an insulating material with a wide bandgap of approximately 5.0–6.0 eV, as determined through various {\it ab initio} calculation techniques \cite{Radin-13,Radin-12,Garcia-Lastra-13}. Substantial research efforts \cite{Radin-13,Garcia-Lastra-13,Ping-12,Tian-14} have focused on understanding its transport properties, as the accumulation of Li$_2$O$_2$ near the cathode \cite{Bruce-12} can block charge flow and cause the so-called \emph{sudden death} of the battery.

It is now widely accepted that both electron and hole polarons \cite{Garcia-Lastra-13,Ping-12,Sio-19} form in these systems, and their hopping barriers have been characterized to estimate their mobilities \cite{Garcia-Lastra-13}. However, more advanced simulations of electron and hole dynamics, such as modeling electron tunneling between distant O$^{2-}_{2}$ ions, have yet to be fully developed.

Our objective is to progress toward such simulations using {\it second-principles} methods \cite{Garcia-16}. To achieve this, we aim to characterize and model the system's main electronic bands by developing an efficient set of Wannier functions \cite{Marzari-12}. These functions serve as a basis for accurately describing the band structure while employing a small set of highly localized functions.

The most relevant orbitals for describing the behavior of Li$_2$O$_2$ are often associated with nearly isolated O$_2^{2-}$ ions embedded within a hexagonal lattice that also includes Li$^+$ counter-ions \cite{Garcia-Lastra-13} (see Fig.~\ref{fig:Li2O2_geom}). Previous studies have primarily relied on band structure and density-of-states analyses. However, to the best of our knowledge, there has been no direct characterization of the localization and shape of the corresponding Wannier functions.

Using our parameterized model, we will investigate how the Wannier one-electron Hamiltonian varies with changes in the system's geometry. This approach will provide a chemically intuitive understanding of the strong electron-phonon coupling present in the system, which plays a critical role in polaron formation~\cite{Sio-19}.

The rest of the paper is organized as follows. The method on which the simulations are based is
described in Sec.~\ref{sec:methodology}. 
In Sec.~\ref{sec:structural}, we discuss the details of the
atomic structure of bulk Li$_{2}$O$_{2}$.
The electronic structure is presented in Sec.~\ref{sec:electronic-structure}, where we also analyze the density of states.
Finally, in Sec.~\ref{sec:wannierization}, we study the Wannier functions and how the Hamiltonian matrix elements expressed in this basis change with the atomic geometry. 

\section{Methodology}
\label{sec:methodology}

Our calculations have been performed within density functional theory~\cite{Hohenberg-64} (DFT) and the generalized gradient approximation (GGA).
We used a numerical atomic orbital (NAO) method, as it is implemented in the SIESTA code~\cite{Soler-02,Garcia-20}.
The exchange-correlation functional was approximated using the Perdew-Burke-Ernzerhof (PBE) functional~\cite{Perdew-96}, as implemented in the {\sc libxc} library~\cite{marques2012libxc,lehtola2018recent}. 

Core electrons were replaced by {\it ab initio} norm-conserving fully separable pseudopotentials~\cite{Kleinman-82}.
In this work the optimized norm-conserving Vanderbilt pseudopotentials proposed by Hamann~\cite{norm_conserving} were used,
in the {\sc psml} format~\cite{psml} available in the Pseudo-Dojo periodic table~\cite{pseudodojo,footnotepseudo}.
For Li, the semicore $1s$ electrons were explicitly included in the valence.
For O, the valence configuration was made of the $2s$ and $2p$ orbitals.

The one-electron Kohn-Sham eigenstates were expanded in a basis of strictly localized~\cite{Sankey-89} numerical atomic orbitals~\cite{Artacho-99}.
Basis functions were obtained by finding the eigenfunctions
of the isolated atoms confined within the soft-confinement spherical potential proposed in Ref.~\cite{Junquera-01}.
A single-$\zeta$ basis set was applied to the $1s$ semicore states of Li. For the valence states of Li and O, we used basis sets of varying sizes, ranging from double-$\zeta$ to triple-$\zeta$, corresponding to two or three radial functions per occupied valence angular momentum shell in the free atom ($2s$ for Li, and $2s$ and $2p$ for O). To enhance angular flexibility, higher angular momentum polarization orbitals (not occupied in the free atom) were included, with an additional shell of $2p$ orbitals for Li and $3d$ orbitals for O, using one (single-polarized) or two (double-polarized) radial functions per polarization shell. All parameters defining the basis functions for Li and O were optimized variationally at the relaxed structure obtained with a converged plane-wave code, following the method in Ref.~\cite{Junquera-01}.

The electronic density, Hartree, and exchange-correlation potentials, as well as the corresponding matrix elements between the basis orbitals, were calculated in a uniform real space grid~\cite{Soler-02}. An equivalent plane-wave cutoff of 600 Ry was used to represent the charge density. 
The integrals in reciprocal space were well converged, using in all the cases a sampling in reciprocal space of the same quality as the ($8\times 8 \times 4$) Monkhorst-Pack mesh~\cite{Monkhorst-76}.

Atomic coordinates were relaxed using a conjugate gradient algorithm until the maximum component of the force on any atom was smaller than 10 meV/$\rm{\AA}$, and the maximum component of the stress tensor was below 0.0001 eV/$\rm{\AA}^{3}$. 

For a given functional and pseudopotential, the converged-basis limit is achieved by a plane-wave calculation with a very high cutoff. To assess the convergence of our NAO basis, we compared the results from {\sc siesta} with those obtained using {\sc abinit}~\cite{gonze2009abinit,gonze2016,gonze2020}.
We aimed to keep the simulations as comparable as possible. Both codes can read identical pseudopotentials in the {\sc psml} format, using the same decomposition into a local pseudopotential operator and Kleinman-Bylander projectors. 
They also share the same exchange-correlation functional, drawn from the same version of the {\sc libxc} library, as well as an identical $k$-point sampling quality and Fermi-Dirac occupation function. The only variation lies in the plane-wave basis set, for which a converged cutoff of 50 Ha was chosen.

\section{Structural properties}
\label{sec:structural}

Li$_{2}$O$_{2}$ has traditionally been associated with two different structures (both of them hexagonal) experimentally measured in the 1950s: the one proposed by Feh\'er~\cite{Feher-53} (belonging to symmetry group P-6, 174) and the one proposed by F\"oppl~\cite{Foppl-57} (belonging to the P6$_{3}$/mmc space group, 194). Later experimental studies, such as the one presented in Ref.~\cite{Chan-11}, using a combination of X-ray and first-principles simulations, showed that the structure proposed by F\"oppl~\cite{Foppl-57} is the most suitable for Li$_{2}$O$_{2}$. 
This was also supported by previous first-principles density functional theory simulations~\cite{Cota-05}, where symmetrized structures for both configurations were compared.
Therefore, the F\"oppl structure within the P6$_{3}$/mmc symmetry group has been used to date to conduct first-principles studies on excitonic effects and their relationship with vibronic coupling~\cite{Garcia-Lastra-11}, as well as polaron dynamics~\cite{Garcia-Lastra-13,Sio-19}.

\begin{table}
    \centering
    \caption{\justifying Atomic positions of the symmetry inequivalent atoms of bulk Li$_{2}$O$_{2}$ in the  hexagonal P6$_{3}$/mmc space group.
             }
    \label{table:wyckoff}
    \begin{tabular}{ccccc}  
    \hline
    \hline
    Wyckoff       &
    Element       &
    $x$           &
    $y$           &
    $z$           \\
    \hline
    $2a$          &
    Li            &
    0             &
    0             &
    0             \\
    $2c$          &
    Li            &
    1/3           &
    2/3           &
    1/4           \\
    $4f$          &
    O             &
    1/3           &
    2/3           &
    $z$ (O)       \\
    \hline
    \hline
    \end{tabular}
\end{table}

\begin{figure}
    \centering
    \includegraphics[width=\linewidth]{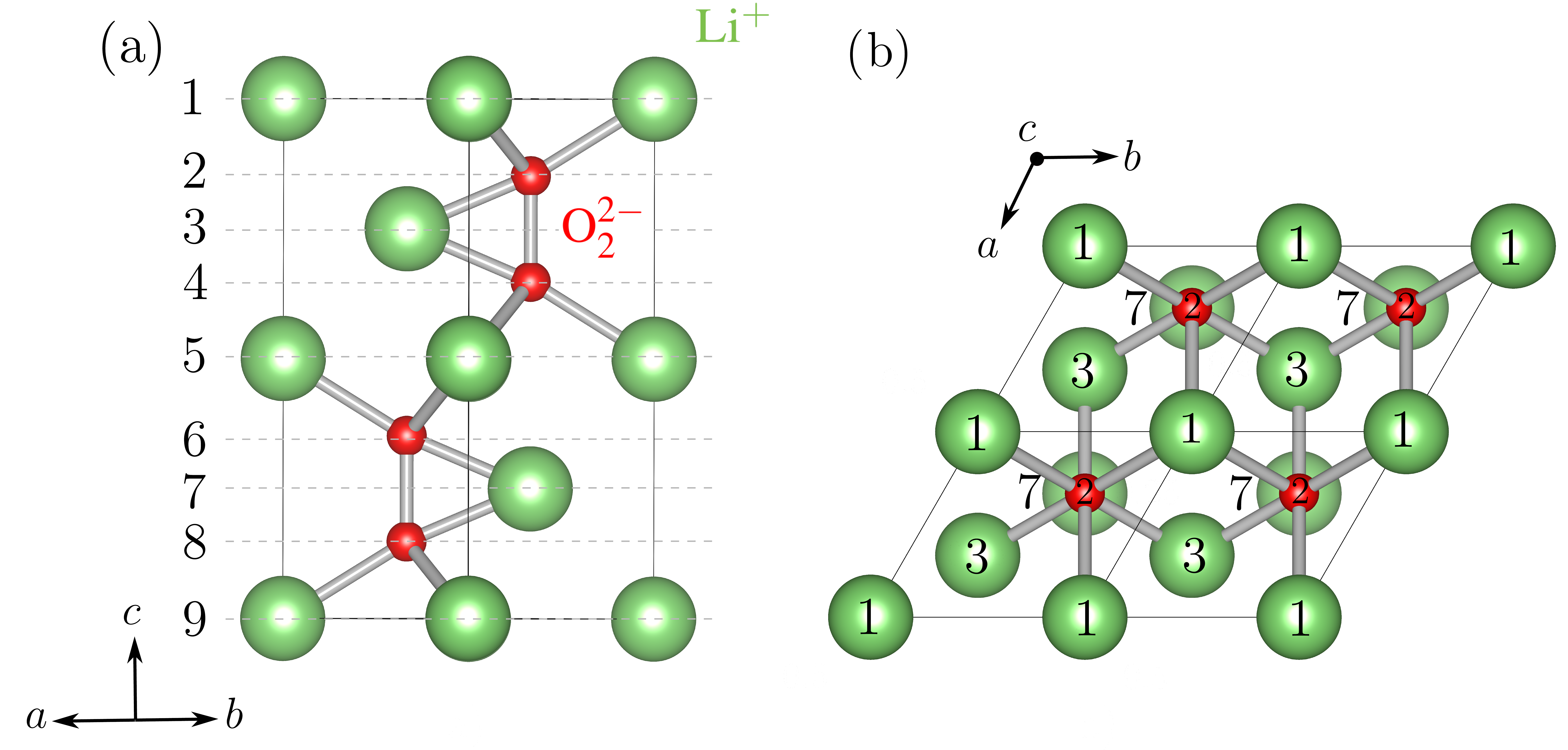}
    \caption{\justifying Schematic representation of the structure of bulk Li$_{2}$O$_{2}$ with the P6$_3$/mmc symmetry (F\"oppl’s structure,~\cite{Foppl-57}). (a) Lateral view. Numbers at the left represent the ordering of the layers along the $z$-direction. (b) Top view. Numbers written on the spheres make reference to the layer that a given atom occupies, according to the ordering given in panel (a).
    Li (respectively O) atoms are represented by green (respectively red) spheres. }
    \label{fig:Li2O2_geom}
\end{figure}

In F\"oppl’s revised structure for Li$_2$O$_2$ with P6$_3$/mmc symmetry, schematized in Fig.~\ref{fig:Li2O2_geom}, the unit cell contains two formula units, with lithium atoms positioned between adjacent oxygen planes along the $c$-axis. The oxygen atoms are arranged into two O$^{2-}_2$ dimers, each oriented parallel to the $c$-axis and separated by lithium layers. Within this structure, the lithium ions are symmetrically coordinated between the O$^{2-}_2$ dimers, stabilizing the crystal in a layered arrangement where each dimer is aligned perpendicular to the basal plane, forming a hexagonal close-packed framework characteristic of the P6$_3$/mmc space group. 
The symmetry inequivalent Wyckoff positions are summarized in Table~\ref{table:wyckoff}. 

Table~\ref{table:structure} illustrates the convergence of NAO basis sets for bulk Li$_{2}$O$_{2}$ by comparing results obtained with different optimized basis sizes. These results are benchmarked against the converged plane-wave (PW) calculations at a 50 Ha cutoff, representing the converged-basis limit, while keeping all other calculation parameters identical.
It is essential to distinguish this converged-basis limit from PW calculations performed at lower cutoffs, which are commonly used in many studies. The converged PW calculations, where the primary sources of error stem from the exchange-correlation functional and the pseudopotentials, slightly overestimate the lattice parameters, with deviations of approximately 0.5\% for $a$ and 0.4\% for $c$, which are typical for the PBE functional.
Although the convergence of NAO basis sets is not systematically achieved by simply increasing the basis size, the sequence of bases in Table~\ref{table:structure} demonstrates clear convergence trends for the in-plane lattice constant and the internal coordinate of the oxygen atom controlling the oxygen-oxygen distance inside the O$_2^{2-}$ dimers.

\begin{table}
    \centering
    \caption{\justifying Structural properties of bulk Li$_{2}$O$_{2}$
             in the hexagonal P6$_{3}$/mmc space group.
             $a$, $b$, and $c$ refers to the length of the three lattice vectors of the conventional unit cell (in \AA ).
             $\alpha$, $\beta$, and $\gamma$ represent the 
             angles between the three lattice vectors.
             $z$(O) stands for the $z$-coordinate of the O
             atom in the Wyckoff positions, according to 
             Table~\ref{table:wyckoff}, given in reduced coordinates.
             $d_{\rm{OO}}$ is the oxygen-oxygen distance
             inside the O$_2^{2-}$ dimers (in \AA ).
             DZP, TZP, and TZDP stands for double-$\zeta$ polarized, triple-$\zeta$ polarized, and triple-$\zeta$ double polarized, respectively.
             PW stands for a plane wave calculation
             carried out with the {\sc abinit} code, with a cutoff of 50 Ha.
            The experimental geometry is taken 
            from Ref.~\cite{Foppl-57}
             }
    \label{table:structure}
    \begin{tabular}{ccccccccc}      
    \hline
    \hline
    Basis set     &
    $a$           &
    $b$           &
    $c$           &
    $\alpha$      &
    $\beta$       &
    $\gamma$      &
    $z$ (O)       &
    $d_{\rm{OO}}$  \\
    \hline
    DZP           &
    3.1593        &
    3.1593        &
    7.6651        &
    90$^\circ$    &
    90$^\circ$    &
    120$^\circ$   &
    0.6484        &
    1.557\\
    TZP             &
     3.1589         &
     3.1589         &
     7.6670         &
    90$^\circ$      &
    90$^\circ$      &
    120$^\circ$     &
    0.6488          &
    1.552           \\
    TZDP            &
    3.1582          &
    3.1582          &
    7.6610          &
    90$^\circ$      &
    90$^\circ$      &
    120$^\circ$     &
    0.6488          &
    1.550           \\
    PW              &
    3.1579          &
    3.1579          &
    7.6840          &
    90$^\circ$      &
    90$^\circ$      &
    120$^\circ$     &
    0.6496          &
    1.543           \\
    Expt.           &
    3.1420          &
    3.1420          &
    7.6500          &
    90$^\circ$      &
    90$^\circ$      &
    120$^\circ$     &
     0.6510         &
     1.515          \\
    \hline
    \hline
    \end{tabular}
\end{table}

\section{Electronic structure}
\label{sec:electronic-structure}

\begin{figure*}[!t]
    \centering
    \begin{subfigure}[b]{0.32\textwidth}    
        \includegraphics[width=\textwidth]{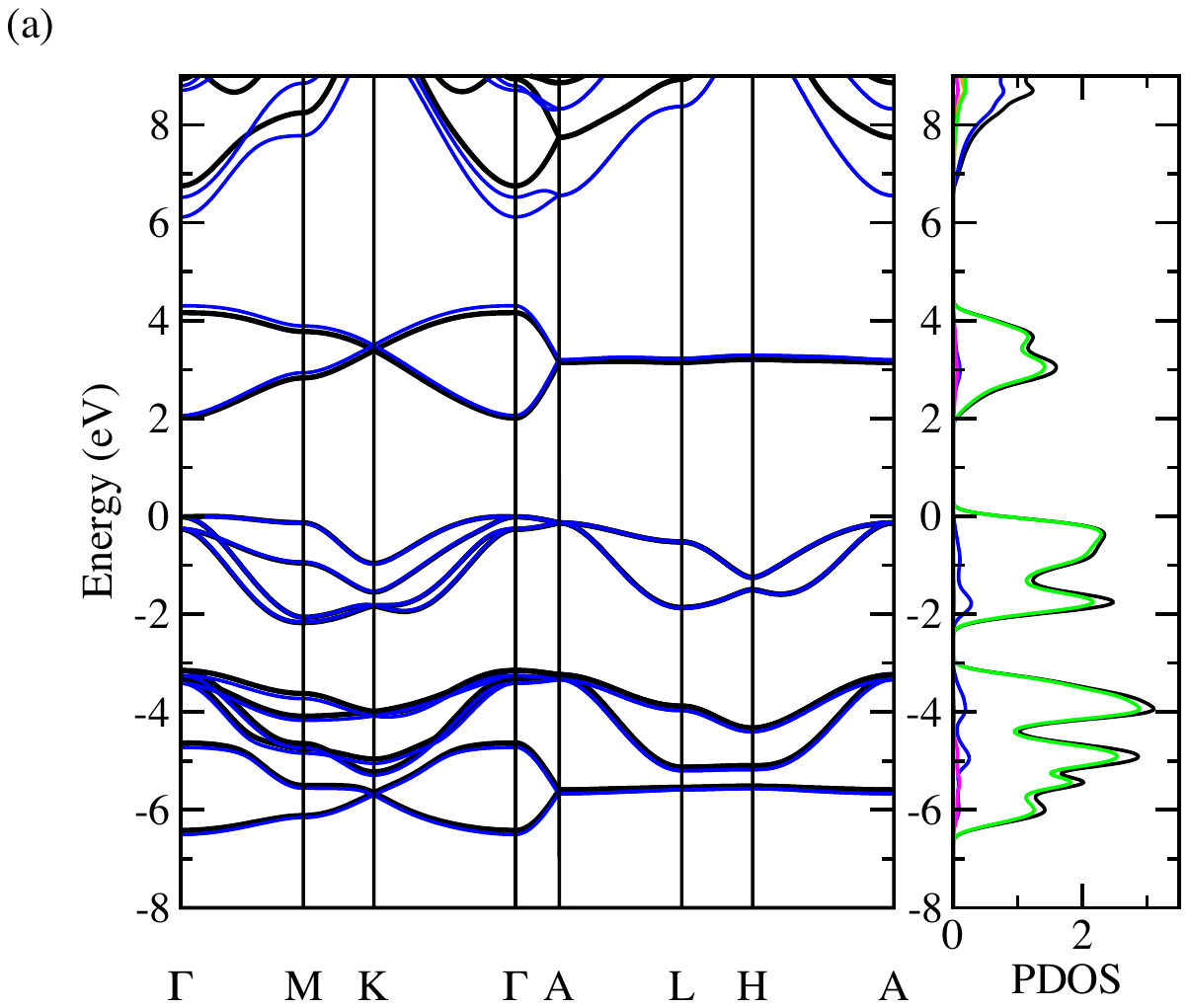}
        \label{fig:figure1}
    \end{subfigure}
    \begin{subfigure}[b]{0.32\textwidth}    
        \includegraphics[width=\textwidth]{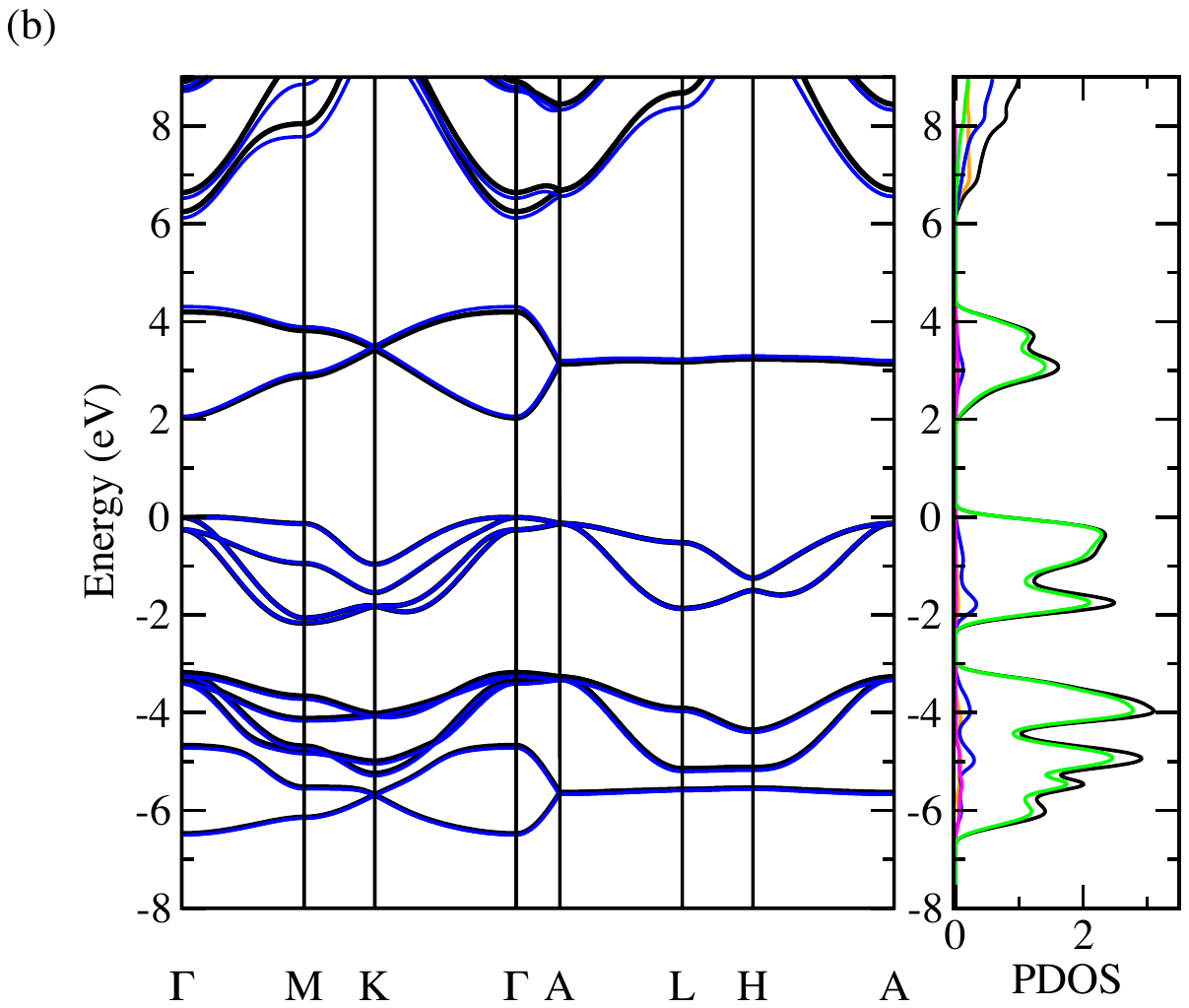}
        \label{fig:figure2}
    \end{subfigure}
    \begin{subfigure}[b]{0.32\textwidth}    
        \includegraphics[width=\textwidth]{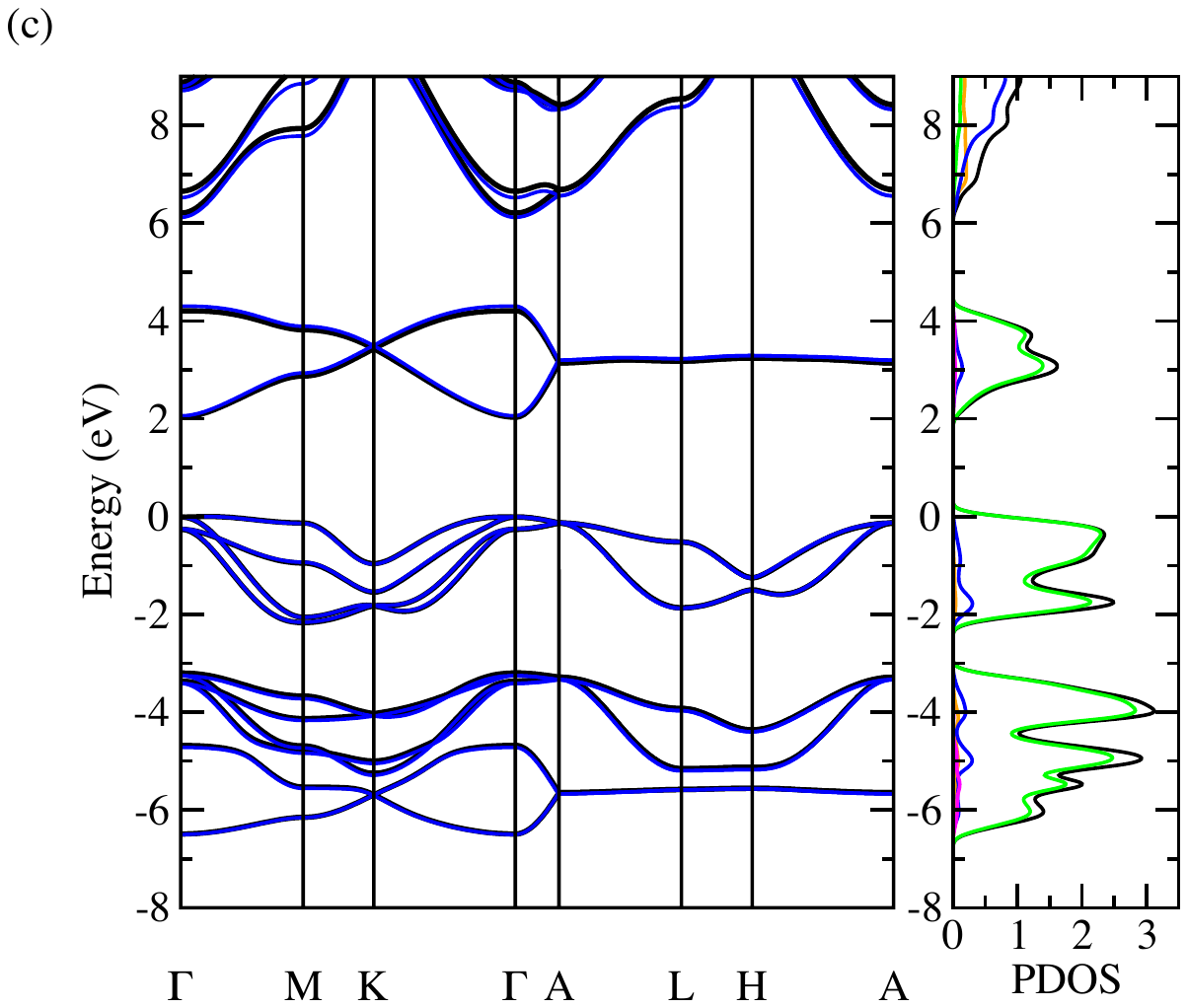}
        \label{fig:figure3}
    \end{subfigure}
    \caption{\justifying Left panels: Electronic band structure of bulk Li$_{2}$O$_{2}$ with P6$_{3}$/mmc symmetry at the relaxed structure (see Table ~\ref{table:structure}) for various basis set sizes. Black lines show the bands calculated with {\sc siesta} for (a) DZP, (b) TZP, and (c) TZDP basis sets. Blue lines correspond to the bands obtained using {\sc abinit} with a converged plane wave cutoff of 50 Ha. The energy zero point is set to the top of the valence band in each case. The projected density of states is shown to the right of each band structure plot, with black lines indicating the total density of states. The projections onto the Li $2s$, Li $2p$, O $2s$, and O $2p$ orbitals are represented by orange, blue, magenta, and green lines, respectively.
    }
    \label{fig:band-structure}
\end{figure*}

The electronic band structure of bulk Li$_{2}$O$_{2}$ in the P6$_{3}$/mmc symmetry at the relaxed structure for different basis set sizes is presented in Fig.~\ref{fig:band-structure}. This structure is consistent with that of a predominantly ionic material. The band manifolds are largely well-separated and each has a distinct dominant character. In this system, the top of the valence band and the bottom of the conduction band are primarily of O $2p$ character, representing the highest occupied energy levels and the lowest unoccupied levels of an oxygen dimer.

\begin{figure}[t!]
\centering
\includegraphics[width=\columnwidth]{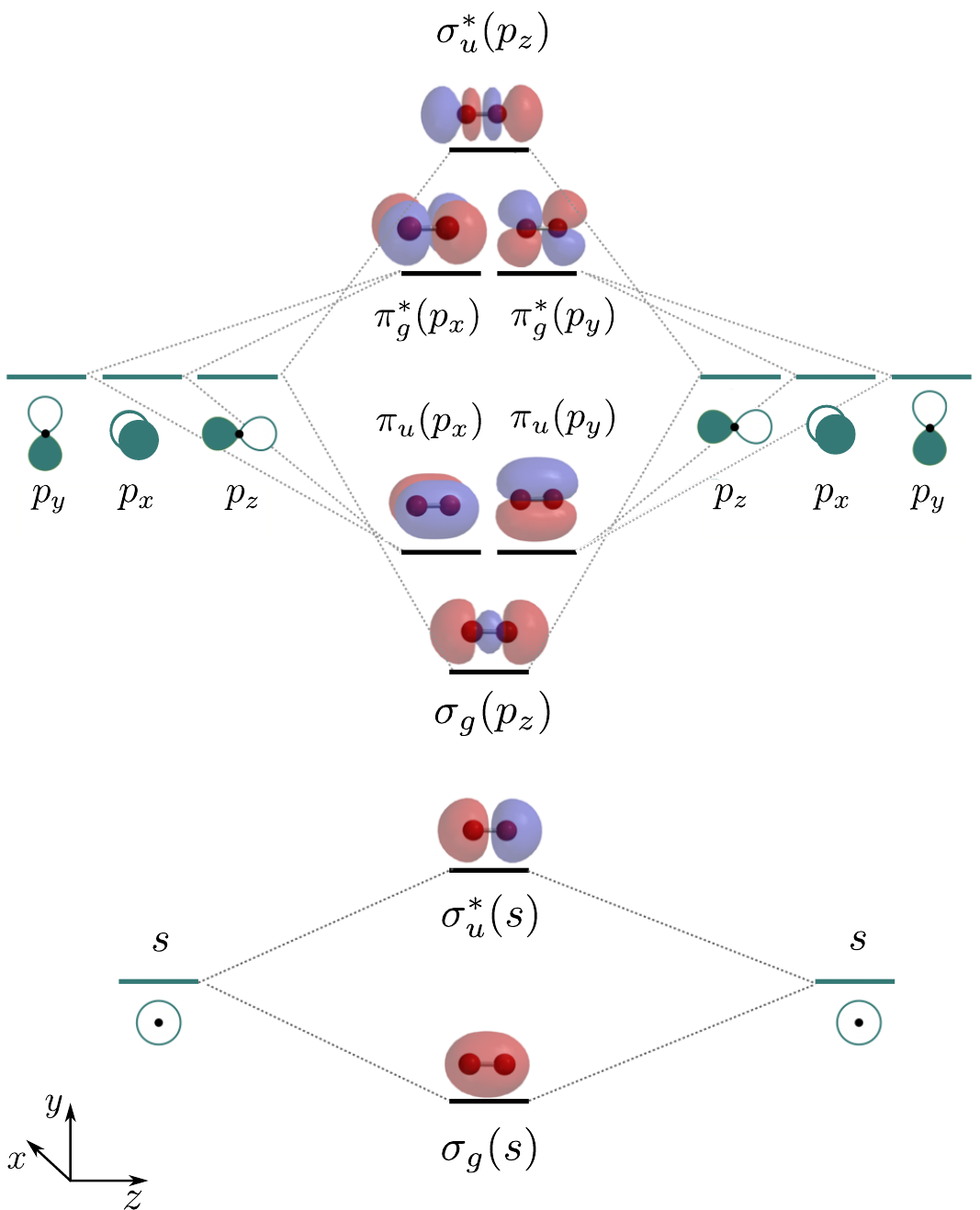}
\caption{\justifying Diagram of the molecular orbitals of the O$_{2}$ dimer and the atomic orbitals from which they are derived.
}
\label{fig:molorb}
\end{figure}

Indeed, the band structure depicted in Fig.~\ref{fig:band-structure} strongly resembles that of the molecular orbitals of the oxygen dimer.
In a purely ionic model, the two Li atoms transfer their $2s$ electrons to the oxygen atoms, resulting in the formation of a ${\rm O}_{2}^{2-}$ peroxide ion. The atomic orbitals of the oxygen atoms then hybridize to create the molecular orbitals of the dimer, as illustrated in Fig.~\ref{fig:molorb}.
The $p_{x}$, $p_{y}$, and $p_{z}$ orbitals of the two oxygen atoms combine to produce six molecular orbitals, ordered by increasing energy as follows:
(i) A bonding $\sigma_{g}$ orbital,
\begin{equation}
    \sigma_{g} = \frac{1}{\sqrt{2}} 
    \left( p_{z}^{(1)} - p_{z}^{(2)}  \right),
\end{equation}

\noindent a symmetric linear combination of the $p_{z}$ orbitals.
(ii) Two degenerate bonding $\pi_{u}$ orbitals,

\begin{equation}
    \pi_{u}[y] = \frac{1}{\sqrt{2}} 
    \left( p_{y}^{(1)} + p_{y}^{(2)}  \right),
\end{equation}

\noindent symmetric combinations of the $p_{y}$ (or $p_{x}$) orbitals.
(iii) Two degenerate antibonding $\pi^{\ast}_{g}$ orbitals,

\begin{equation}
    \pi^{\ast}_{g}[y] = \frac{1}{\sqrt{2}} 
    \left( p_{y}^{(1)} - p_{y}^{(2)}  \right),
\end{equation}

\noindent antisymmetric combinations of the $p_{y}$ (or $p_{x}$) orbitals.
(iv) An antibonding $\sigma^{\ast}_{u}$ orbital,

\begin{equation}
    \sigma^{\ast}_{u} = \frac{1}{\sqrt{2}} 
    \left( p_{z}^{(1)} + p_{z}^{(2)}  \right),
\end{equation}

\noindent an antisymmetric linear combination of the $p_{z}$ orbitals.

Each oxygen atom contributes with four $p$ electrons, which, combined with the two electrons transferred from the Li atoms, totals ten electrons. These electrons occupy the five lowest-energy molecular orbitals, filling them completely (below the Fermi energy).
Since this is a closed shell structure, no spin-polarized calculations are justified. 

The analogy between bands and molecular orbitals would be established as follows: first, the molecular orbital $\sigma_{g}$ corresponds to the band manifold between -6 and -4 eV, the degenerate $\pi_{u}$ orbitals map to the band manifold spanning -5 to -3 eV, and the degenerate $\pi^{\ast}_{g}$ orbitals align with the top of the valence band manifold. The unoccupied sixth molecular orbital, $\sigma^\ast_{u}$, corresponds to the lowest conduction band manifold.

The overall agreement between the {\sc siesta} and {\sc abinit} bands is excellent, particularly for the valence bands and the first conduction band manifold, where differences are minimal. 
With a DZP-quality basis [Fig.~\ref{fig:band-structure}(a)], slight discrepancies appear in the conduction band manifold with Li $2p$ character, but these differences are resolved when the basis set quality is increased to TZP [Fig.~\ref{fig:band-structure}(b)] or TZDP [Fig.~\ref{fig:band-structure}(c)].

\begin{figure}[t!]
\centering
\includegraphics[width=0.8\columnwidth]{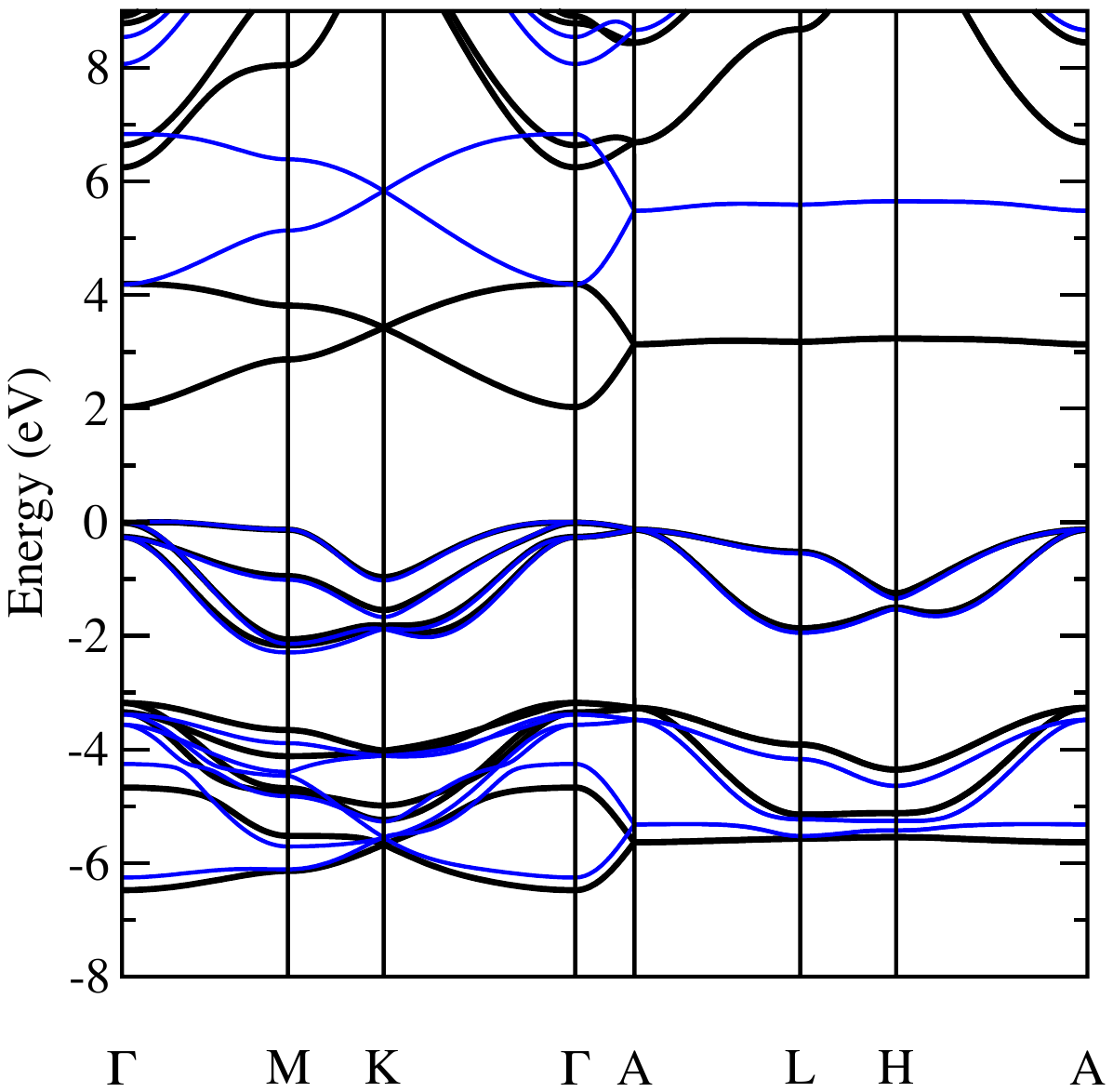}
\caption{\justifying Electronic band structure of bulk Li$_{2}$O$_{2}$ with P6$_{3}$/mmc symmetry obtained with {\sc siesta} with the PBE functional (black lines) and the HSE06 hybrid functional with a percentage of exact exchange of 25 \%.
The quality of the basis set has been fixed to TZP.
}
\label{fig:hybrids}
\end{figure}

At energies above 6 eV from the top of the valence band, the bands shift to a dominant Li $2p$ character. The computed band gap, using the PBE functional, is direct at the $\Gamma$ point and measures 2.0 eV.
This value is not directly comparable to the experimental value which, to our knowledge has not been measured, due to the well-known DFT band gap misfit.
The experimental optical spectrum shows an absorption onset around 3.3 eV although this low-energy features are associated to excitonic phenomena, as proven by Bethe-Salpeter simulations~\cite{Garcia-Lastra-11}.
A more accurate estimation of the band gap itself is obtained by an average of G$_0$W$_0$ and self-consistent GW calculations~\cite{Radin-13} that yields a value of 6.73 eV.
In Fig.~\ref{fig:hybrids} we show the Li$_2$O$_2$ band structure as calculated with the HSE06 functional~\cite{HSE06}, recently implemented in {\sc siesta}~\cite{Garcia-20}.
We can see that, while the qualitative shape of the bands is very similar to those obtained with PBE, Fig.~\ref{fig:band-structure}, the band gap is significantly increased to 4.18 eV although it is still smaller than the reference 6.73 eV, which can be obtained when the Hartree-Fock mixing parameter is increased from the 0.25 value in HSE06 to 0.48~\cite{Radin-13}.

\section{Wannierization}
\label{sec:wannierization}

\subsection{Projection on atomic-like orbitals}

In the previous section, we calculated the band structure of bulk Li$_2$O$_2$ using three optimized basis sets: DZP, TZP, and TZDP, which correspond to 76, 96, and 128 orbitals per unit cell, respectively. Diagonalizing the Kohn-Sham Hamiltonian yields the same number of bands per $k$-point in the first Brillouin zone, spanning from deeply localized semicore Li orbitals to numerous conduction band orbitals.

In many studies, it is advantageous to focus on the electronic states at the top of the valence band and the bottom of the conduction band, as these states are most relevant for capturing the underlying physics. Such an approach is particularly useful for investigating excitonic effects~\cite{Garcia-Lastra-11} or electron and hole polarons~\cite{Garcia-Lastra-13}. Here, this method provides insights into the strong electron-vibration coupling within the O$_2^{2-}$ ions.

A powerful way to address this problem is to construct a basis of localized, orthogonal Wannier functions~\cite{Marzari-97, Marzari-12}, which offer a minimal and efficient representation of the bands of interest. These functions not only simplify the Hamiltonian for electronic structure analysis but also enable second-principles methods~\cite{Garcia-16}. The {\sc scale-up} code incorporates these methods, leveraging symmetry to compute electron-lattice and electron-electron corrections within the simplified Hamiltonian in the Wannier basis. To ensure model accuracy, it is crucial for the Wannier basis to preserve the system's symmetry.

For this reason, we chose ``maximally projected Wannier functions'' rather than the conventional maximally localized Wannier functions~\cite{Marzari-97}. These functions are created by projecting Bloch states onto atomic orbitals (used as initial guess functions) without minimizing the spread functional, as implemented in the {\sc wannier90} code~\cite{Pizzi-20}. Given that both the valence band maximum and conduction band minimum are dominated by O $2p$ character (see Fig.~\ref{fig:band-structure}), it is logical to use the twelve $p$-type atomic orbitals in the unit cell (three per oxygen atom for the four oxygen atoms).

\begin{figure}[H]
    \centering
    \includegraphics[width=\linewidth]{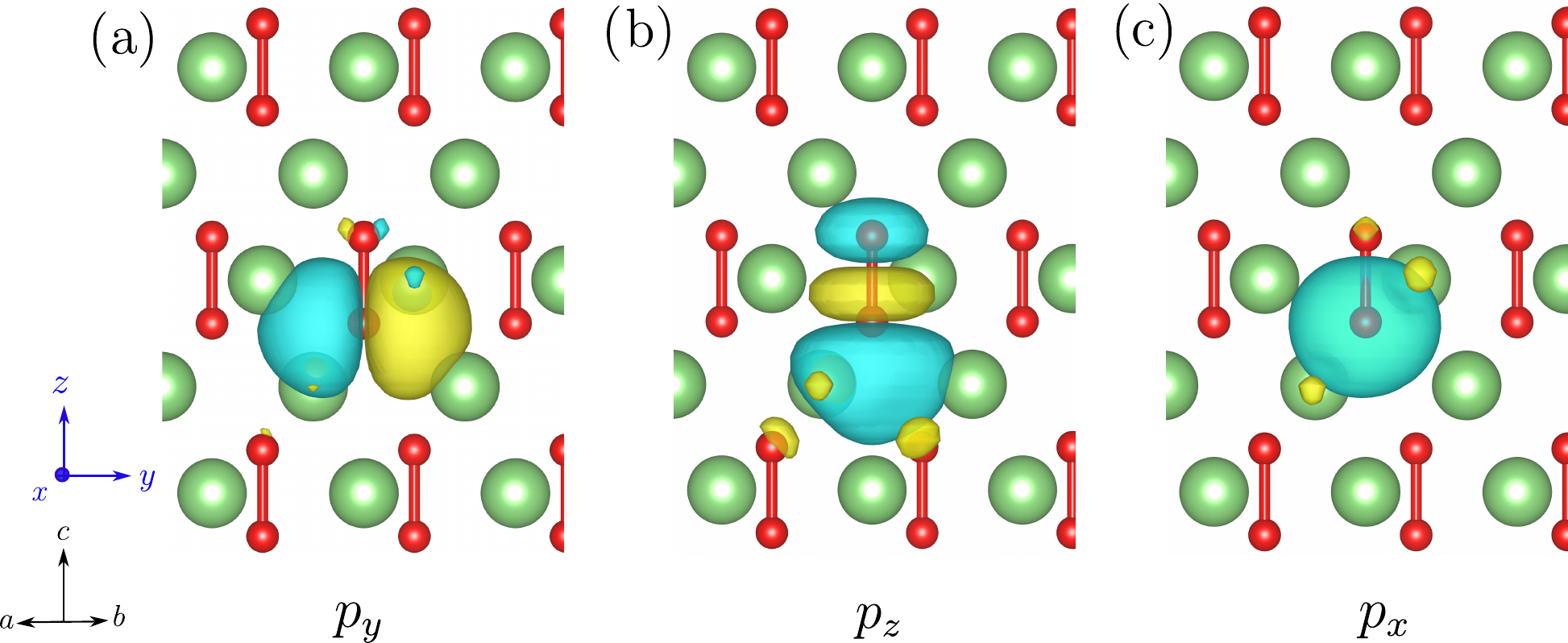}
    \caption{\justifying Representation of Wannier functions obtained by projecting Bloch orbitals onto the numerical basis orbitals of {\sc siesta} for the Oxygen atom with $p$-symmetry: (a) for $p_{y}$, 
    (b) for $p_{z}$, and (c) for $p_{x}$. Meaning of the balls as in Fig.~\ref{fig:Li2O2_geom}.}
    \label{fig:Wanniers}
\end{figure}

Figure~\ref{fig:Wanniers} shows the Wannier functions derived using this approach. The $p_x$ and $p_y$ orbitals, which are symmetry-equivalent in the P$6_3$/mmc space group, closely resemble $p$-orbitals of a hydrogen-like atom, except for small nodes near neighboring Li atoms, ensuring orthogonality with adjacent Wannier functions.

In contrast, the $p_{z}$-like functions, which align along the O-O bond direction in the ${\rm O}_{2}^{2-}$ dimers, exhibit a significantly different chemical environment. 
Here, the changes in the localized functions due to the inter-orbital orthogonality condition are larger, due to the stronger overlap between $p_z$ functions and, as a result, they no longer resemble atomic orbitals as closely.
These Wannier orbitals, while showing the characteristic node of the $2p$ orbitals on the atom they are centered on, also display a lobe over the opposite O-atom in the dimer.  The outer lobe, pointing away from the O$_2^{2-}$ dimer, is larger than the inner one, possibly reflecting on small orbital contributions to the wavefunction from the surrounding Li$^+$. 
The orbitals in this quasi-atomic representation are very compact with spreads of 0.75 and 0.85 \AA$^2$ for $\pi$($p_x$,$p_y$) and $\sigma$($p_z$) orbitals, respectively. 
We have checked that these values are quite stable and do not change significantly with the quality of the basis set employed during their calculation. 

\subsection{Projection on molecular orbitals}
\label{sec:wannier-mol}

\begin{figure}[H]
    \centering
    \includegraphics[width=\linewidth]{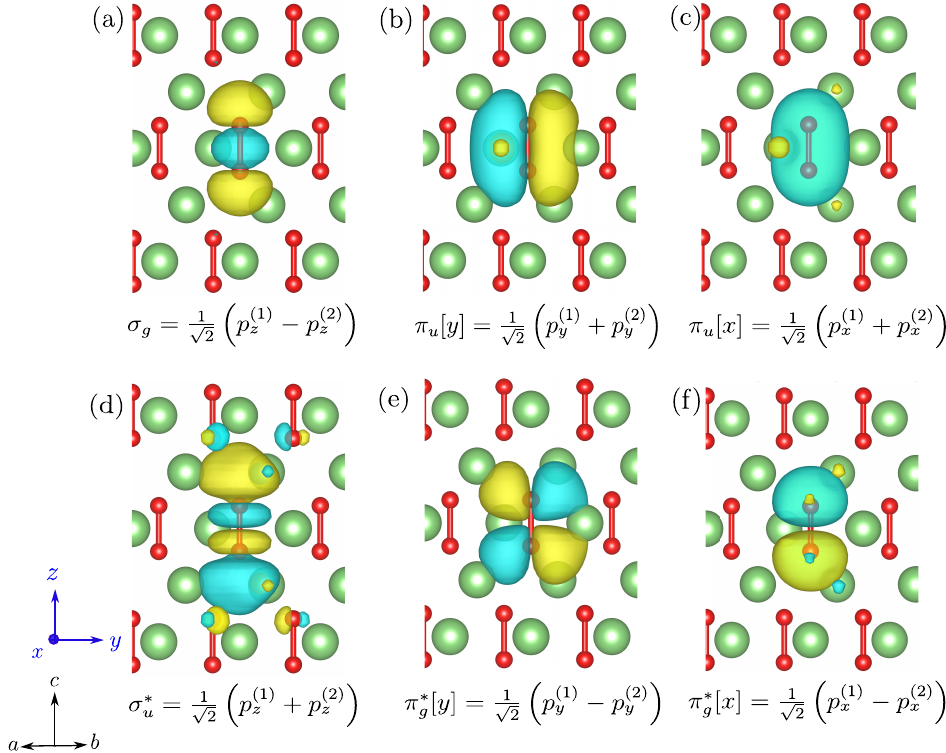}
    \caption{\justifying Representation of Wannier functions associated to bands with marked molecular-orbital character. The (a)-(f) panels correspond, respectively, with Wannier orbitals with strong $\sigma_g(p_z)$, $\pi_u(p_y)$, $\pi_u(p_x)$, $\sigma_u^\ast(p_z)$, $\pi_g^\ast(p_y)$ and $\pi_g^\ast(p_x)$  character. }
    \label{fig:mol-orbitals}
\end{figure}

In the previous localization scheme all the Bloch bands with a majority weight on the oxygen $2p$ bands were mixed together in the unitary transformation that yields to the Wannier functions to provide the most compact orbitals that describe these bands. 
However, this method also forces the use of all oxygen bands (twelve functions per primitive cell), and all their inter-orbital interactions, when creating a reduced model. 
Given that the bands of Fig.~\ref{fig:band-structure} associated to each of the molecular orbitals of the O$_2^{2-}$ ion do not cross each other they can be Wannierized individually.
This means that the bands with $\sigma_g$ (two), $\pi_u$ (four), $\pi_g^\ast$ (four), and $\sigma_u^\ast$ (two) character can each be treated separately, reducing the number of bands in the electronic structure model and significantly limiting the number of inter-orbital matrix elements. For example, a Wannier model for electron polarons would require only $\sigma_u^\ast$ bands, while a model for hole polarons would require only $\pi_g^\ast$ bands.
The resulting localized orbitals, illustrated in Fig.~\ref{fig:mol-orbitals}, have larger spreads than quasi-atomic orbitals: 1.60, 1.29, 1.49, and 2.49 \AA$^2$ for $\sigma_g$, $\pi_u$, $\pi_g^\ast$, and $\sigma_u^\ast$ bands, respectively. Bonding $\pi$-orbitals are more compact than both bonding and antibonding $\sigma$-orbitals, as well as antibonding $\pi$-orbitals. Among these, the antibonding $\sigma$-orbitals are particularly diffuse.
Comparing panels (a) and (f) in Fig.~\ref{fig:mol-orbitals} we can see that the main difference is that for the Wanniers coming from the $\sigma_{u}^\ast$ bands, outer lobes of the antibonding orbital are expanded with respect to the former towards the Li$^+$ ions. 
This is a sensible result since these unoccupied orbitals are energetically close to the more dispersive band with Li character and take on some of their character. 
Thus we see that, while the character of the bands clearly corresponds to the molecular orbital diagram shown in Fig.~\ref{fig:molorb}, there are some differences with the idealized orbitals that are important to highlight. 
For example, close inspection of the $\pi$-type orbitals [Fig.~\ref{fig:mol-orbitals}(b)-\ref{fig:mol-orbitals}(e)], shows deformation from the ideal cylindrical-symmetry shape of a 2$p$ orbital related to the presence of nearby Li$^+$ ions.
Thus, while the main character of these orbitals is clearly molecular, their shape is clearly influenced by the embedding of the O$_2^{2-}$ ions in the solid and these small changes can, in turn, be important to describe other phenomena.

\subsection{Electron-vibration coupling}

\begin{figure*}[!t]
    \centering
    \includegraphics[width=\linewidth]{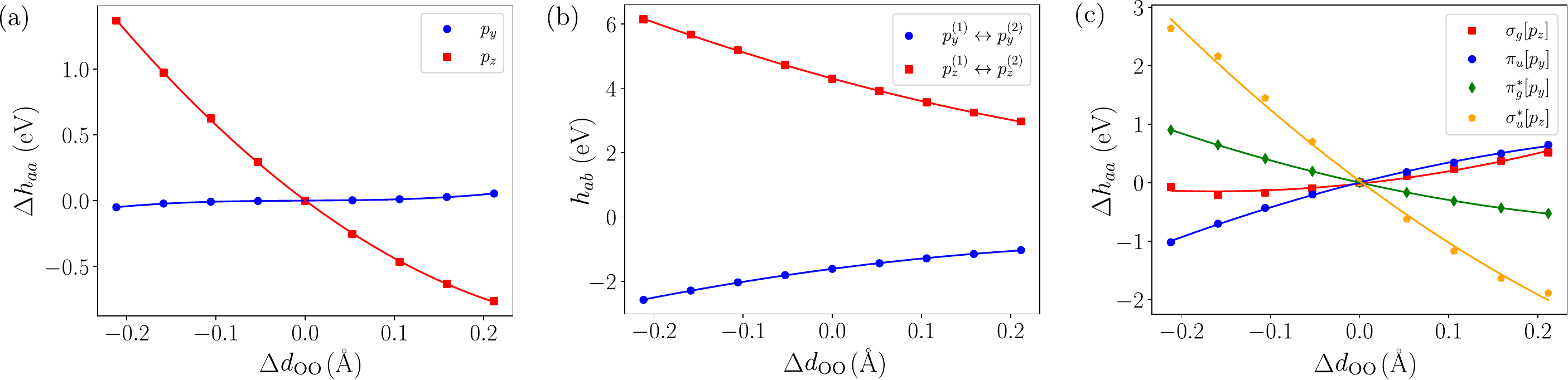}
    \caption{\justifying Variation of the hamiltonian matrix elements in the Wannier basis with the change of oxygen-oxygen distance, $\Delta d_{\rm{OO}}$, inside the O$_2^{2-}$ dimer. In (a) the change of self-energy ($\Delta h_{aa}$) of quasi-atomic $p_z$ (red) and $p_x/p_y$ orbitals (blue) with respect to the equilibrium position is presented. In (b), for the same basis, the interaction matrix element, $h_{ab}$, between $p_z$($p_x/p_y$) orbitals in the same O$_2^{2-}$ unit is presented in red(blue). Finally, in (c) the change of self-energy for Wannier functions with strong $\sigma_g$, $\pi_u$, $\pi_g^\ast$, and $\sigma_u^\ast$-like character with respect to the equilibrium distance is presented in red, blue, green and yellow, respectively. }
    \label{fig:Wannier_elvib}
\end{figure*}

We now turn our attention to examining how the one-electron Hamiltonian varies with the oxygen-oxygen distance for the two Wannier function families discussed in the previous section.
This is illustrated in Fig.~\ref{fig:Wannier_elvib}, where panels (a) and (b) show the changes in the diagonal ($h_{aa}$) and off-diagonal ($h_{ab}$) elements, respectively, for quasi-atomic Wannier functions, while panel (c) presents the variation of the diagonal matrix elements for molecular-type Wannier orbitals.
Figure~\ref{fig:Wannier_elvib}(a) shows that the self-energy of quasi-atomic Wannier orbitals with $p_z$ character exhibits a strong dependence on the oxygen-oxygen distance in a dimer ($d_{\rm{OO}}$), whereas the self-energy of the degenerate $p_x$ and $p_y$ orbitals shows a much weaker variation.
This is likely due to the fact that $p_x$ and $p_y$-like orbitals are localized around only one atom of the dimer, whereas $p_z$-like orbitals have significant contributions centered on both atoms in the dimer (see Fig.~\ref{fig:Wanniers}) and exhibit rapid changes as the distance varies.
Moreover, in the quasi-atomic representation, $p_z$ orbitals primarily interact with each other to form the $\sigma$ bonding and antibonding bands, while $p_x$ and $p_y$ orbitals interact similarly to generate the $\pi$ bands.
These interactions give rise to new off-diagonal $h_{ab}$ parameters, representing the one-electron hamiltonian between Wannier functions a and b, which necessitate the simultaneous representation of both $\sigma$ ($\pi$) bonding and antibonding bands, making it impossible to separate them within this basis.
At the equilibrium position, the $p_z$ interaction element is significantly larger (4.30 eV) than the $p_x$/$p_y$ element (-1.62 eV), consistent with the stronger $\sigma$ interactions compared to $\pi$ interactions.
The variation of these elements with $d_{\rm{OO}}$ is shown in Fig.~\ref{fig:Wannier_elvib}(b), with red and blue representing the $p_z$ and $p_x$/$p_y$ orbitals, respectively.
It can be observed that as the distance increases, the absolute value of the off-diagonal matrix elements decreases in both cases, ultimately approaching each other in the limit of infinite distance, where the interaction between the orbitals on each atom vanishes.
Finally, Fig.~\ref{fig:Wannier_elvib}(c) shows the variation of the self-energies in the molecular-Wannier basis, as described in Sec.~\ref{sec:wannier-mol}.
In this figure, we observe that the energies of the bonding orbitals (represented in red and blue for $\sigma$ and $\pi$ orbitals, respectively) decrease as the oxygen atoms move closer, while the opposite trend is seen for the antibonding orbitals (depicted in green and yellow for $\pi$ and $\sigma$ orbitals, respectively).
Although one might initially expect the energy of the $\sigma$ orbitals to vary more rapidly than that of the $\pi$ orbitals, this is only true for the antibonding $\sigma$ level. The bonding $\sigma$ orbital, in contrast, exhibits a much slower variation.
This slower variation arises from the 2$s$-2$p$ hybridization, which affects the ordering of the $\sigma_g(p_z)$ and $\pi_u(p_x,p_y)$ molecular orbitals in homonuclear diatomic molecules across the series C$_2$, N$_2$, O$_2$, and F$_2$ \cite{Atkins_PhysChem}. This hybridization shifts much of the expected rapid variation of the primarily $\sigma_g(p_z)$ orbital to the deeper $\sigma_g$ orbital with significant 2$s$ character.
We also find, as expected, that only self-energies need to be described when describing the hamiltonian using a molecular-like basis. 

Ultimately, we can observe that describing the variation of most Wannier Hamiltonian elements with a second-order polynomial, as shown in Fig.~\ref{fig:Wannier_elvib}, produces accurate results.
The curve where the error is larger is that corresponding to the $\sigma^\ast_u$ orbital where the changes in energy are larger. 
While the fit with a second-order polynomial clearly captures the order of magnitude and main tendencies of the curve, high-accuracy in the range -0.2 to 0.2 \AA\ can only be achieved using a third-order polynomial.
This is highly promising for developing models that incorporate precise electron-vibration interactions in this system.
Notably, the results presented here clarify why, in previous calculations of the hole polaron \cite{Garcia-Lastra-13}, the O-O bond distance decreases—due to the removal of an antibonding electron—while for the electron polaron, where an electron is added to the $\sigma$ antibonding orbital, the O-O bond distance increases significantly.
Thus, adopting a molecularly inspired Wannier basis appears to be the most suitable approach for constructing reduced models to simulate polaron motion dynamics on a large scale using {\it second-principles}.
This basis is smaller and more flexible, focusing on describing only the HOMO- and LUMO-type bands (or either one), and its parameters offer a clearer and simpler physical interpretation compared to those of quasi-atomic Wannier functions. While quasi-atomic Wannier functions can sometimes be more compact, the molecularly inspired basis provides distinct advantages in interpretability and adaptability.

\section{Conclusions}

In this work, we have comprehensively investigated the structural and electronic properties of bulk lithium peroxide, a key material in lithium-air battery technologies. Our study leveraged density functional theory and second-principles approaches to provide insights into the atomic structure, electronic band structure, and electron-lattice coupling in this material.

The structural analysis was carried out for the P$6_{3}$/mmc hexagonal symmetry, that provides a stable framework for Li$_{2}$O$_{2}$, characterized by well-defined oxygen dimers and symmetrically coordinated lithium ions. 
Electronic structure calculations revealed that the valence and conduction bands are dominated by oxygen 2$p$ states, which exhibit clear correspondence to the molecular orbitals of the peroxide ion. 
The results obtained with a basis set of numerical atomic orbitals of triple-zeta polarized quality present good agreement with those obtained with a converged plane-wave basis set. The calculated band gap, while underestimated using standard DFT, aligns qualitatively with the known insulating nature of Li$_{2}$O$_{2}$.

Our Wannierization approach successfully localized the electronic states into both atomic-like and molecular-like Wannier functions, with the latter offering a more chemically intuitive description of the bands relevant for polaron formation and dynamics. By analyzing the variation of Hamiltonian elements in the Wannier basis as a function of geometry, we demonstrated the strong electron-phonon coupling in the system and its implications for the behavior of polarons. Notably, the molecular Wannier basis provided an efficient and physically interpretable framework, particularly for understanding the structural changes associated with electron and hole polarons.

The findings presented here emphasize the utility of molecularly inspired Wannier functions for constructing reduced models that accurately capture the dynamics of polarons and electron-vibration interactions in Li$_{2}$O$_{2}$. Such models are crucial for large-scale simulations of charge transport and could inform the design of next-generation lithium-air batteries.

\section{Acknowledgements}
P.M.M., E.J.T., P.V.K.R, B. M., and V. S. ~acknowledge financial support from Erasmus+ KA-107 action and the Vice-rectorate for Internationalisation and Global Engagement of the University of Cantabria.
T.F.R., N.C.S, P.G.F, and J.J.~acknowledge financial support from Grant No.~PID2022-139776NB-C63 funded by MCIN/AEI/10.13039/501100011033 and by ERDF ``A way of making Europe'' by the European Union.
T.F.R. acknowledges financial support from Ministerio de Ciencia, Innovaci\'on y Universidades (Grant PRE2019-089054).
N.C.S. acknowledges financial support from ``Concepci\'on Arenal'' Grant No. BDNS:524538 of the University of Cantabria  funded by the Government of Cantabria.
%


\begin{thebibliography}{39}
\bibitem [{foo()}]{footnotepseudo}%
  \BibitemOpen
  \href@noop {} {}\bibinfo {note} {The scalar relativistic \textsc{oncvpsp} v0.4.1 pseudopotentials with stringent accuracy were used.}\BibitemShut {Stop}%
\end{thebibliography}

%

\end{document}